\documentstyle[11pt,newpasp,twoside,epsf]{article}
\def\edcomment#1{\iffalse\marginpar{\raggedright\sl#1\/}\else\relax\fi}
\reversemarginpar

\begin{document}
\title{Globular Clusters as Probes of the Virgo gE NGC 4472}
\author{Terry Bridges$^1$, Keith Ashman$^2$, Mike Beasley$^3$, Doug Geisler$^4$,
Dave Hanes$^5$, Ray Sharples$^3$, \& Steve Zepf$^6$}
\affil{Anglo-Australian Observatory$^1$, University of Kansas$^2$, University of Durham$^3$,
Concepcion University$^4$, Queen's University$^5$,
Yale$^6$}

\begin{abstract}
We present radial velocities for 144 globular clusters (GCs) around the 
Virgo gE NGC 4472 (M49), and ages and metallicities from co-added
spectra of 131 GCs.  We confirm our earlier finding that the
metal-poor GCs have a significantly higher velocity dispersion than the
metal-rich GCs, and we find little or no rotation in the
metal-rich GCs.  The velocity dispersion profile is consistent
with isotropic orbits for the GCs and the NGC 4472
mass distribution inferred from X-ray data.  Our sample
of GCs spans a metallicity range of $-1.6 \leq [Fe/H] \leq 0$ dex.
The metal-poor and metal-rich GC populations are coeval within
the errors, and all GCs older than 6 Gyr at 95\% confidence.
Our findings are consistent 
with a merger origin for NGC 4472, but other elliptical formation models
cannot be ruled out. 
\end{abstract}

\section{Observations}

GC spectra were obtained in two observing runs, one in 1994
with WHT/LDSS-2 (Sharples et al. 1998), 
and the second in 1998 with CFHT/MOS (Beasley et al. 2000; Zepf
et al. 2000).
GC candidates for both runs were chosen from Washington photometry
(Geisler et al. 1996), with 
colour ($0.5 < C-T_1 < 2.2$) and magnitude ($19.5 < V < 22.5$) selection.
Spectra were obtained
over the wavelength range 3600$-$6000 \AA, with a resolution of 
3$-$6 \AA, and a velocity precision of 50$-$100 km/sec.  
Exposure times were 3$-$3.5 hours per mask.  
We have a total sample of 144 confirmed GCs in NGC 4472,
out to $\sim$ 7\arcmin~ radius ($\sim$ 30 kpc, 6 R$_{eff}$).  See
Sharples et al., and Beasley et al. for more details.

\section{GC Kinematics and Dark Matter in NGC 4472 (Zepf. et al. 2000)}

{\bf Figure 1} shows smoothed velocity and velocity dispersion
profiles 
for the full GC sample,
and for the metal-rich and metal-poor populations separately.
We confirm our earlier result (Sharples et al. 1998) that the metal-poor
GCs have a significantly higher velocity dispersion than the 
metal-rich GCs.
There is little or no 
rotation in the metal-rich GCs, and modest rotation of $\sim$ 100
km/sec in the metal-poor GCs.  
For the metal-rich GCs, V/$\sigma$ $<$ 0.34 at 99\%
confidence.  This absence of rotation in the more
centrally-concentrated metal-rich GCs seems to require
significant outward angular momentum transport (from a merger?). 

\begin{figure}
\plotfiddle{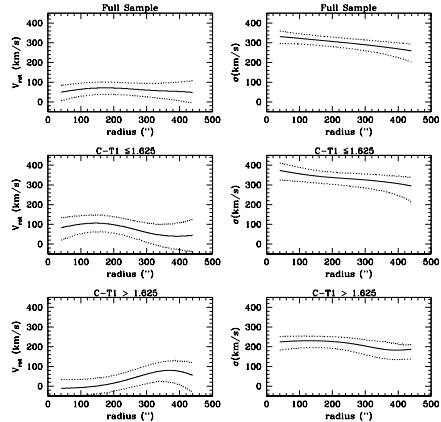}{1.75in}{0}{30}{30}{-100}{-60}
\caption{Smoothed rotation and velocity dispersion profiles 
for NGC 4472 GCs.  {\bf Top:}~  
full dataset; {\bf Middle:}~ metal-poor (blue) GCs;
{\bf Bottom:}~ metal-rich (red) GCs.  Dotted lines show the
1$\sigma$ uncertainties, determined from bootstrapping.}
\end{figure}

We have used the deprojected GC velocity dispersion and
density profiles together with the Jeans equation to derive
the mass distribution of NGC 4472, with the assumption of isotropic
GC orbits. 
{\bf Figure 2} shows the resulting
mass profile for NGC 4472, compared with that obtained from X-ray
data (Irwin \& Sarazin 1996).  There is reasonable
agreement between the two profiles, 
and we confirm the
existence of a substantial dark matter halo in NGC 4472, with a
M/L ratio $> 50$ at 30 kpc radius.

\begin{figure}
\plotfiddle{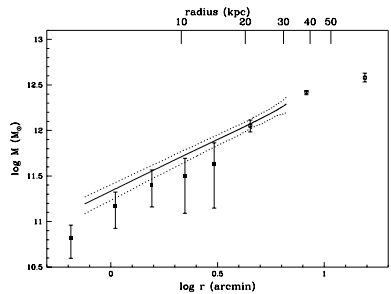}{1.25in}{0}{30}{30}{-100}{-85}
\caption{The mass of NGC 4472 as a function of radius.  The 
solid line is the best fit to 144 GC radial velocities,
and the dotted lines are the 1$\sigma$ lower and upper limits.  The
points are masses inferred from ROSAT X-ray data (Irwin \& Sarazin 1996).}
\end{figure}

\section{GC Ages and Abundances (Beasley et al. 2000)}

To improve the S/N ratio for age and abundance analysis, 
we have co-added our GC spectra by colour into four bins with 30$-$35
GCs each.  For each bin, we have measured age (H$\beta$, H$\delta$, H$\gamma$)
and metallicity (Mg$_2$, Fe5270/5335) sensitive indices,
on the Lick system.  {\bf Figure 3} shows 
the predictions of Worthey (1994) models for the NGC 4472 GCs in 
the Mg$_2$, H$\beta$ plane.  
Based on a Galactic GC calibration, 
the NGC 4472 GCs span a metallicity
range  
$-1.6 \leq [Fe/H] \leq 0$ dex. 

It is notoriously difficult to determine {\it absolute} GC ages,
but within the errors all four bins are coeval; we can further say
that all GCs are older than 6 Gyr at 95\% confidence.  
Interestingly, the metal-rich GCs are 0.7 $\pm$ 7 Gyr younger
than the metal-poor GCs.  {\bf Table 1} summarizes the ages and
metallicities for our binned NGC 4472 data.

\begin{figure}
\plotfiddle{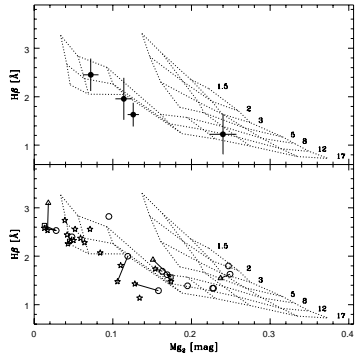}{1.5in}{0}{25}{25}{-100}{-50}
\caption{Predictions of Worthey (1994) models for NGC 4472
and Galactic GCs.  {\bf Top Panel:}~ co-added
NGC 4472 GCs (filled circles).
{\bf Lower Panel:}~Data for Galactic GCs. Open circles:~
Cohen et al. (1998); open stars:~Brodie \& Huchra (1990);
triangles:~our CFHT data; squares:~our WHT data.  Lines
connect the same GCs between different datasets.  Ages on the right
side of the grid run from 1.5$-$17 Gyr.}
\end{figure}

\begin{table}
\caption{Ages and abundances for binned NGC 4472 GCs.} 
\begin{tabular}{cccc}
\tableline
Bin \# & $\overline{C-T_1}$ & [Fe/H] (dex) & Mean Age (Gyr) \\
\tableline
1 & 1.30 $\pm$ 0.09 & $-$1.29 $\pm$ 0.30 & 10.7 $^{+6}_{-5}$\\
2 & 1.44 $\pm$ 0.07 & $-$0.91 $\pm$ 0.35 & 15.3 $^{+8}_{-5}$\\
3 & 1.61 $\pm$ 0.06 & $-$0.84 $\pm$ 0.25 & 18.5 $^{+4}_{-5}$\\
4 & 1.91 $\pm$ 0.11 & $-$0.27 $\pm$ 0.30 & 11.3 $^{+8}_{-9}$\\
\tableline
\tableline
\end{tabular}
\end{table}

\noindent{\bf References}

\medskip

\parindent 0pt

Beasley, M., et al. 2000, MNRAS, 318, 1249 

Geisler, D., Lee, M.G., \& Kim, E. 1996, AJ, 111, 1529

Irwin, J.A., \& Sarazin, C.L. 1996, ApJ, 471, 683

Sharples, R.M., et al. 1998, AJ, 115, 2337

Worthey, G. 1994, ApJS, 95, 107

Zepf, S.E., et al. 2000, AJ, in press (astroph/0009130)


\end{document}